\documentclass[pra, aps,twocolumn,showpacs]{revtex4}
\usepackage{bm, epsfig}
\usepackage{amsmath}
\usepackage{graphicx, subfigure}
\begin{document}
\title{Bichromatic control of dynamical tunneling: influence of the irregular Floquet states}
\author{Archana Shukla}
\affiliation{Department of Chemistry, Indian Institute of Technology,
Kanpur U. P. 208 016, India}
\author{Srihari Keshavamurthy}
\affiliation{Department of Chemistry, Indian Institute of Technology,
Kanpur U. P. 208 016, India}
\date{\today}

\begin{abstract}
Bichromatic control, in terms of the amplitude and relative phase of the second field as control knobs, is an useful approach for controlling a variety of quantum processes. In this context, understanding the features of the control landscape is important to assess the extent and efficiency of the control process. A key question is whether, for a given quantum process, one can have regions wherein there is a complete lack of control. In this work we show that such regions do exist and can be explained on the basis of the phase space nature of the quantum Floquet states. Specifically, we show that robust regions of no control arise due to the phenomenon of chaos-assisted tunneling. We also comment on the possible influence of such regions on the phenomenon of directed transport in quantum Hamiltonian ratchets.

\end{abstract}

\maketitle

\section{Introduction}
Coherent control of quantum tunneling is currently an area of active research.  In many instances one is faced with the task of controlling dynamical tunneling - a process wherein the barriers are in the phase space and the system tunnels between two symmetrically related regular regions in the phase space\cite{hellerdavis81,heller95}. Dynamical tunneling is rather ubiquitous in nature and  manifests in systems ranging from the dynamics of trapped cold atoms\cite{hensinger01,steck01,lenz13} to the  flow of  vibrational energy in molecules\cite{ksjcp05,kspre05,ksirpc}. Over the last decade, considerable advances\cite{dyntunbook} have been made in our understanding of the mechanism of dynamical tunneling. It is now well established that dynamical tunneling is extremely sensitive to the classical phase space structures. In particular, for systems in the mixed regular-chaotic regimes, the role of the nonlinear resonances resulting in resonance-assisted tunneling (RAT)\cite{brodier01,eltschka05,mouchet06} and the influence of the chaotic sea leading to the phenomenon of chaos-assisted tunneling (CAT)\cite{bohigas93,tomsovic94} have been studied in great detail. Clearly, any attempt to control dynamical tunneling must therefore take into cognizance the interplay of RAT and CAT occurring in the system.

An attractive possibility, given that the tunneling happens between symmetry related regions, is to break the symmetry and arrest the tunneling dynamics. A natural approach when dealing with periodically driven systems is the so called laser induced symmetry breaking\cite{franco06,franco08,franco10}. Thus, consider a Hamiltonian periodically driven by a field with amplitude $E_{1}$ and frequency $\omega$
\begin{eqnarray}
H_{1}(q,p,t) &=& H_{0}(q,p) - \mu(q) {\cal E}_{1}(t) \\
&\equiv& H_{0}(q,p) -E_{1} \mu(q) \cos(\omega t). \nonumber
\label{onefieldham}
\end{eqnarray}
As usual, it is advantageous to characterize the solutions $\Psi_{\alpha}(q,t) = e^{-i \epsilon_{\alpha}t/\hbar} \Phi_{\alpha}(q,t)$ to the time-dependent Schr\"{o}dinger equation in terms of the time periodic Floquet eigenstates $\Phi_{\alpha}(q,t) = \Phi_{\alpha}(q,t+2\pi/\omega)$ and the associated quasienergies $\epsilon_{\alpha}$.
If the above Hamiltonian possesses the dynamical symmetry\cite{flach00,franco08}
\begin{equation}
H_{1}\left(-q,-p,t+\frac{\pi}{\omega}\right) = H_{1}(q,p,t)
\label{symma}
\end{equation}
then the Floquet states of eq.~\ref{onefieldham} come as even-odd symmetric doublets $\Phi_{\alpha}^{\pm}(q,t)$. The quasienergy splitting of the doublets $\Delta \epsilon_{\alpha} = |\epsilon_{\alpha}^{+} - \epsilon_{\alpha}^{-}|$ relates to the timescale of dynamical tunneling. 
 In order to break the symmetry and control the tunneling process one introduces a second (control) field to the Hamiltonian in eq.~\ref{onefieldham} as follows
 \begin{equation}
 H(q,p,t) = H_{1}(q,p,t) - E_{2} \mu(q) \cos(2\omega t + \theta),
 \label{twofieldham}
 \end{equation}
 with $E_{2}$ and $\theta$ being the control field amplitude and relative phase respectively. 
For $E_{1},E_{2} \neq 0$ the symmetry of eq.~\ref{symma} is violated. Moreover, for the relative phase $\theta \neq 0,\pm \pi$ an additional symmetry $(q,p,t) \rightarrow (q,-p.-t)$ is also violated. Incidentally, note that the violation of the two symmetries is a prerequisite for observing directed transport {\it i.e.,} ratcheting in driven Hamiltonian systems\cite{flach00,denisov01,denisov07,denisov14}. Bichromatic control Hamiltonians as in eq.~\ref{twofieldham} have proved to be useful in many contexts including reaction dynamics\cite{shapiro88,constantoudis05}, ionization of atoms\cite{schafer92,schumacher94,ehlotzky01,sirko03} and oriented molecules\cite{ohmura06,ohmura04}, control of population imbalance of trapped BEC\cite{molina08}, high harmonic generation\cite{fleischer05,bandrauk05}, and orientation of rotationally cold molecules\cite{kanai01,guerin02,de09,spanner12}.

From the above arguments it appears that controlling dynamical tunneling by bichromatic fields as in eq.~\ref{twofieldham} should be a relatively easy task.  However, there are indications\cite{latka94,sethi08} that this viewpoint is too simplistic in situations when the tunneling dynamics involves, apart from the doublets, additional Floquet states.  For instance, an earlier study\cite{farrelly93} showed that the driven double well system\cite{lin90,lin92} could be controlled using bichromatic fields. In contrast, a subsequent study\cite{sethi08} of the bichromatic control landscape of the driven double well established that involvement of Floquet states delocalized in the stochastic region of the phase space leads to a lack of control of the tunneling dynamics. Nevertheless, there are important questions that still need to be answered. For instance, is there a specific class of delocalized Floquet states that can lead to a complete loss of bichromatic control? What happens to the nature of the bichromatic control landscape upon varying the effective value of the Planck's constant?  In this work we address these questions in detail for a simple yet paradigmatic model of the driven pendulum. We show that the control knobs $(E_{2},\theta)$ are ineffective in certain regimes, further strengthening the notion that symmetry breaking is not always effective in controlling processes dominated by CAT.

\section{Model Hamiltonian and the process of interest}
\label{modham}

The Hamiltonian of interest\cite{denisov07} is a periodically driven pendulum and can be expressed as
\begin{equation}
H(q,p,t) = \frac{1}{2} p^{2} + \left(1 + \cos q \right) - q {\cal E}(t),
\label{ourham}
\end{equation}
where ${\cal E}(t) \equiv E_{1} \cos(\omega t) + E_{2} \cos(2\omega t + \theta)$ is a bichromatic  external  field. 
The driven pendulum is a paradigmatic model system that appears in various contexts. 
Here $E_{1}$ is the amplitude of the driving field and $E_{2}$ and $\theta$ are the amplitude and the relative phase of the controlling field.
Due to the periodicity of the Hamiltonian in eq.~\ref{ourham}, it is advantageous to study the time evolution in terms of the 
Floquet states $|\Phi_{\alpha}\rangle$ and their associated quasienergies 
$\epsilon_{\alpha}$ with $-\pi < \epsilon_{\alpha} < \pi$.

As discussed in the introduction, our interest in this paper is to use the $2\omega$-field to control a dynamical tunneling process occurring in the system for $E_{2} = 0$. We start with a brief overview of the nature of the classical phase space associated with eq.~\ref{ourham} and then specify the quantum process of interest. 

\begin{figure}
\begin{center}
\includegraphics[width=0.5\textwidth]{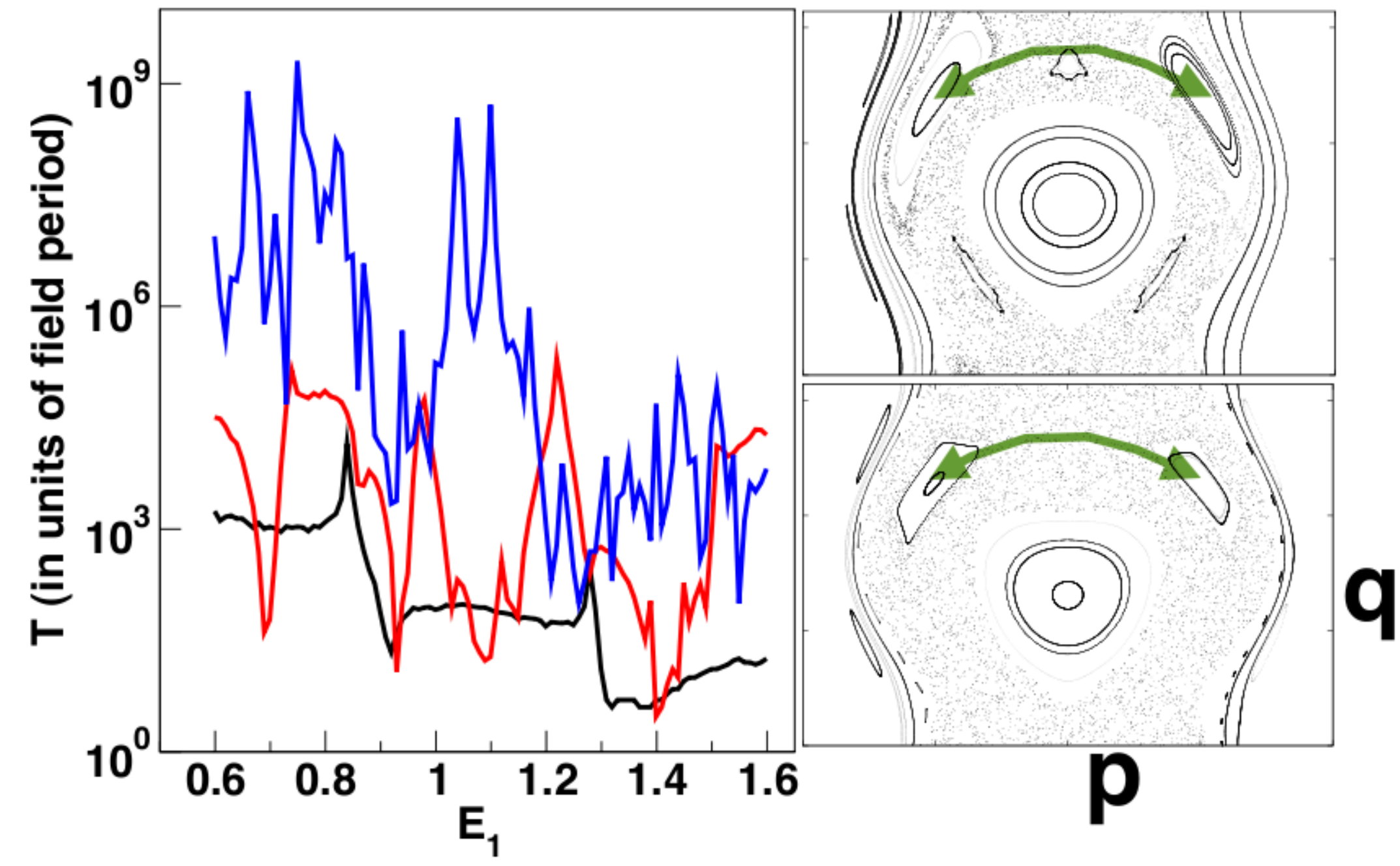}
\caption{The right panels show the stroboscopic surface of section for the driven pendulum in eq.~\ref{ourham} for $E_{1}=0.6$ (top panel) and $E_{1}=1.6$ (bottom panel). The other parameters are $\omega=2$ and $E_{2}=0$. The left panel shows the tunneling (shown as arrows in the phase space) time between the symmetry related $\Omega:\omega = 1:1$ islands for a coherent state placed at the center of the left island. The three data shown in the left panel correspond to $\hbar=0.2$ (black line), $\hbar=0.1$ (red line), and $\hbar=0.05$ (blue line).}
\label{fig1}
\end{center}
\end{figure}

The nature of the phase space for eq.~\ref{ourham}, a $1.5$ degree of freedom system, can be determined using the stroboscopic surface of section. In Fig.~\ref{fig1} the surfaces of section are shown for the range of $E_{1}$ values of interest with $E_{2} = 0$ and the driving field frequency $\omega = 2$. Note that the main features of the phase space are nearly the same throughout the entire range.
In order to identify the various nonlinear resonances one  writes eq.~\ref{ourham} in terms of the action-angle variables $(J,\phi)$ of the unperturbed pendulum Hamiltonian and considering ${\cal E}(t) = E_{1} \cos \omega t$ yields
\begin{equation}
H(J,\phi,t) = H_{0}(J) + E_{1}  \sum_{n=-\infty}^{\infty} V_{n}(J) e^{i(n \phi + m\omega t)},
\end{equation}
with $m = \pm 1$ as the only values. Note that the pendulum action-angle variables are different for trapped ($E_{0} < 2$) and untrapped ($E_{0} > 2$) orbits. Nevertheless, both cases can be written in the above form with different Fourier amplitudes $V_{n}(J)$. In addition, only odd values of $n$ appear for the trapped case. Thus, all field-matter nonlinear resonances are of the form $n \dot{\phi} \equiv n \Omega(J) = \omega$ and are denoted in this work as $n:1$. The nonlinear frequencies can be determined to be
\begin{equation}
\Omega(J) = \begin{cases}
                       \frac{\pi}{2{\cal K}(k_{t})}  \,\,\,\,\,\,\,\,\text{trapped} \\
                       \frac{\pi}{k_{u} {\cal K}(k_{u})}  \,\,\,\,\,\,\,\,\text{untrapped},
                    \end{cases}
\end{equation}                       
with ${\cal K}(k)$ being the complete elliptic integral of the first kind with moduli $k_{t} = \sqrt{E_{0}/2}$ and $k_{u} = \sqrt{2/E_{0}}$.
In Fig.~\ref{fig1} a prominent $\Omega:\omega=1:1$ nonlinear resonance is present. It can be shown that this  corresponds to a resonance between an untrapped (rotor) state of the pendulum and the $\omega$-field. For the values of the effective Planck constant used in this work, this $1:1$ resonance island can support several Floquet states. In addition, a $\Omega:\omega=3:1$ resonance involving a trapped state of the  pendulum and the $\omega$-field can be observed for $E_{1}=0.6$. The $3:1$ resonance, capable of supporting one or two Floquet states and the only prominent resonance for our case, is destroyed for $E_{1}=1.6$.

The main focus of this work is on controlling the dynamical tunneling occurring between the the $\Omega:\omega = 1:1$ resonance islands since the $1:1$ islands play a key role in enhancing the directed current in the system\cite{denisov01,denisov07}. Classically, trajectories in the left island are trapped forever. However, quantum mechanically, it is well known that dynamical tunneling destroys the classical localization. Thus, a coherent state initially localized in the left island of Fig.~\ref{fig1} can tunnel to the symmetry related right island. In Fig.~\ref{fig1} the tunneling time for an initial state
\begin{equation}
\psi(q;q_{c},p_{c}) = \langle q|\psi \rangle = \frac{1}{\left(2 \pi \sigma^{2}\right)^{1/4}} \exp\left[-\frac{(q_{c}-q)^{2}}{4 \sigma^{2}} + i \frac{p_{c}q}{\hbar}\right]
\label{cohstate}
\end{equation}
localized about the island center $(q_{c},p_{c})$ are shown as a function of $E_{1}$ for various values of $\hbar$ with $\sigma^{2} = \hbar/2$.
The quantum dynamics are performed by expressing the initial state as a superposition of the Floquet states
\begin{equation}
|\psi(0)\rangle = \sum_{\alpha} C_{\alpha}(0) |\Phi_{\alpha} \rangle
\end{equation}
and obtaining the time evolved state $|\psi(kT_{f})\rangle$ at integer multiples of the field period $T_{f} = 2 \pi/\omega$ using the Floquet operator.  The Floquet states of eq.~\ref{ourham} are determined using an efficient method used in an earlier work\cite{denisov07}.
The fluctuations, increasing in number and amplitude with decreasing $\hbar$, of the tunneling times seen in Fig.~\ref{fig1} are paradigmatic of the RAT and CAT phenomena. A rather detailed understanding of these fluctuations in terms of the various phase space structures  is possible\cite{ksirpc,mouchet06,bohigas93,lock10}. However, in the present work we are interested in controlling the dynamical tunneling process of Fig.~\ref{fig1} by switching on the second field in eq.~\ref{ourham} and mapping out the bichromatic control landscape.

\section{Bichromatic control landscape}

We now switch on the $2\omega$-field with $\theta \neq 0,\pm \pi$ and study the dynamics of the initial state centered on the left $1:1$ island in Fig.~\ref{fig1}.  As discussed in the introduction, such a symmetry breaking field should destroy the dynamical tunneling process and essentially localize the initial state. In the context of Hamiltonian ratchets, the initial state is centered about a transporting island and symmetry breaking is expected to result in directed motion. In order to obtain a comprehensive view of the control process it is useful to construct a control landscape. 
Such a control landscape should unambiguously show regions of control or lack of control. Several measures can be used to construct the control landscape. For instance, the decay time of the survival probability associated with the initial state
\begin{eqnarray}
S(\tau) &\equiv& |\langle \psi(0)| \psi(\tau) \rangle|^{2} \nonumber \\
&=&\left|\sum_{\alpha} e^{-i\epsilon_{\alpha}\tau/\hbar} \langle  \psi(0)|\Phi_{\alpha}(0)\rangle
\langle \Phi_{\alpha}(0)| \psi(0)\rangle \right|^{2} \nonumber \\
&=& \sum_{\alpha,\beta} p_{\alpha}p_{\beta} e^{-i(\epsilon_{\alpha}-\epsilon_{\beta})\tau/\hbar}
\label{survprob}
\end{eqnarray}
can be used as a measure for the control landscape. In the above, $p_{\alpha} \equiv |\langle  \psi(0)|\Phi_{\alpha}(0)\rangle|^{2}$ is the overlap intensity. Note that the tunneling times shown in Fig.~\ref{fig1} are determined by such an approach. It is also possible to use the time averaged version of eq.~\ref{survprob} as a measure for the control landscape\cite{sethi08}. However, such approaches can become unwieldy for various reasons and in this work we use a simpler measure to map out the landscape. This measure is the long time limit of $S(\tau)$ given by 
\begin{equation}
\lim_{\tau \rightarrow \infty} S(\tau) = \sum_{\alpha} p_{\alpha}^{2} \equiv L(E_{1},E_{2},\theta)
\label{ipr}
\end{equation}
and represents the number of Floquet states that participate in the dynamics of $|\psi(0)\rangle$. In eq.~\ref{ipr} we have denoted  $L(E_{1},E_{2},\theta)$ as the landscape `function'. It is also possible to have the driving frequency $\omega$ as another parameter for the landscape function. However, in case of the present work, this is of limited utility since the nature and location of the $1:1$ island changes with varying $\omega$. The measure in eq.~\ref{ipr} is particularly useful from the dynamical tunneling perspective since CAT and RAT are expected to be multistate processes.

\begin{figure}
\begin{center}
\includegraphics[width=0.45\textwidth]{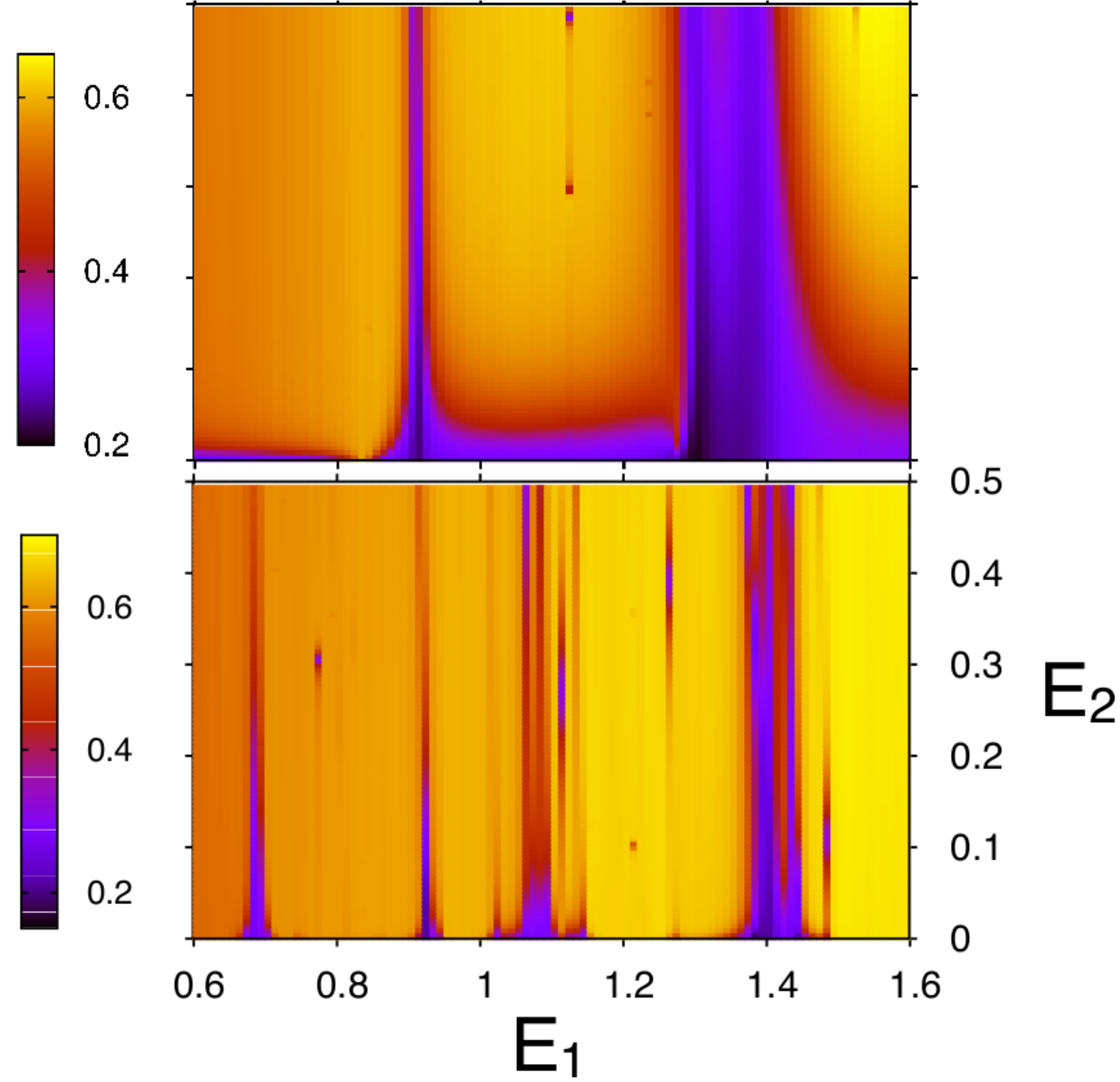}
\caption{Control landscape  $L(E_{1},E_{2},\theta)$ as in eq.~\ref{ipr} for  $\hbar=0.2$ (top) and $\hbar=0.1$ (bottom) with fixed values $\theta = \pi/2$ and $\omega=2$. Lowest values signal multiple states participating in the dynamics of the initial coherent state and lack of bichromatic control. Landscapes computed on an equally spaced $101 \times 101$ grid.}
\label{fig2}
\end{center}
\end{figure}
  
In Fig.~\ref{fig2} we summarize our results for the control landscapes $L(E_{1},E_{2},\pi/2)$ computed using eq.~\ref{ipr}, and  the initial state (cf. eq.~\ref{cohstate}) being localized about the left $1:1$ resonance island in the phase space. In Fig.~\ref{fig2} (top) we show $L(E_{1},E_{2},\pi/2)$  for $\hbar=0.2$ and it is immediately clear that there are distinct regions corresponding to low values of  $L(E_{1},E_{2},\pi/2)$. Our computations (also see Fig.~\ref{fig5}) show that such regions correspond to situations wherein, despite symmetry breaking, the $2\omega$-field  is unable to control the dynamical tunneling process. The calculation for $\hbar=0.1$ case, shown in Fig.~\ref{fig2} (bottom), shows the persistence of some of the no control regions observed  for $\hbar=0.2$. In addition, new regions, for instance around $E_{1} \approx 0.7$, emerge in the control landscape. 

Insights into the origins of such regions of no control seen in Fig.~\ref{fig2} comes from inspecting  the variation of the quasienergy spectrum with $E_{1}$ for $E_{2}=0$ shown in Fig.~\ref{fig3}. Clearly, regions of no control in Fig.~\ref{fig2} correlate well with the  avoided crossings involving the tunneling Floquet doublets with other Floquet states. Although the observation that such avoided crossings can lead to the loss of bichromatic control has been made earlier\cite{latka94}, a comprehensive study of the landscape as a function of the key control knobs $(E_{2},\theta)$ is still lacking. To this end, in Fig.~\ref{fig3} we also show $L(E_{1},E_{2},\theta)$ for fixed values of $E_{1}$ corresponding to three of the avoided crossings. The results show that not all avoided crossings are equally effective in disrupting  bichromatic control.  While there are regions that exhibit a complete loss of control as in Fig.~\ref{fig3}(III), there are regions as in Fig.~\ref{fig3}(II) where control is restored with small variations in the the relative phase $\theta$.   One may argue, and correctly so, that the differences come from the extent of sharpness of the avoided crossings and hence the effective coupling between the Floquet states. Nevertheless, in the following section we show that deeper insights come from investigating the phase space nature of the participating states near the avoided crossings. 

We note that similar correlations between avoided crossings and landscape features, including the mechanisms and arguments presented below, are seen in the $\hbar=0.1$ case as well. Consequently, for the sake of brevity, in what follows we focus on the $\hbar=0.2$ case exclusively.   

\section{Classical-quantum correspondence for bichromatic control}
    
\begin{figure}
\begin{center}
\includegraphics[width=0.5\textwidth]{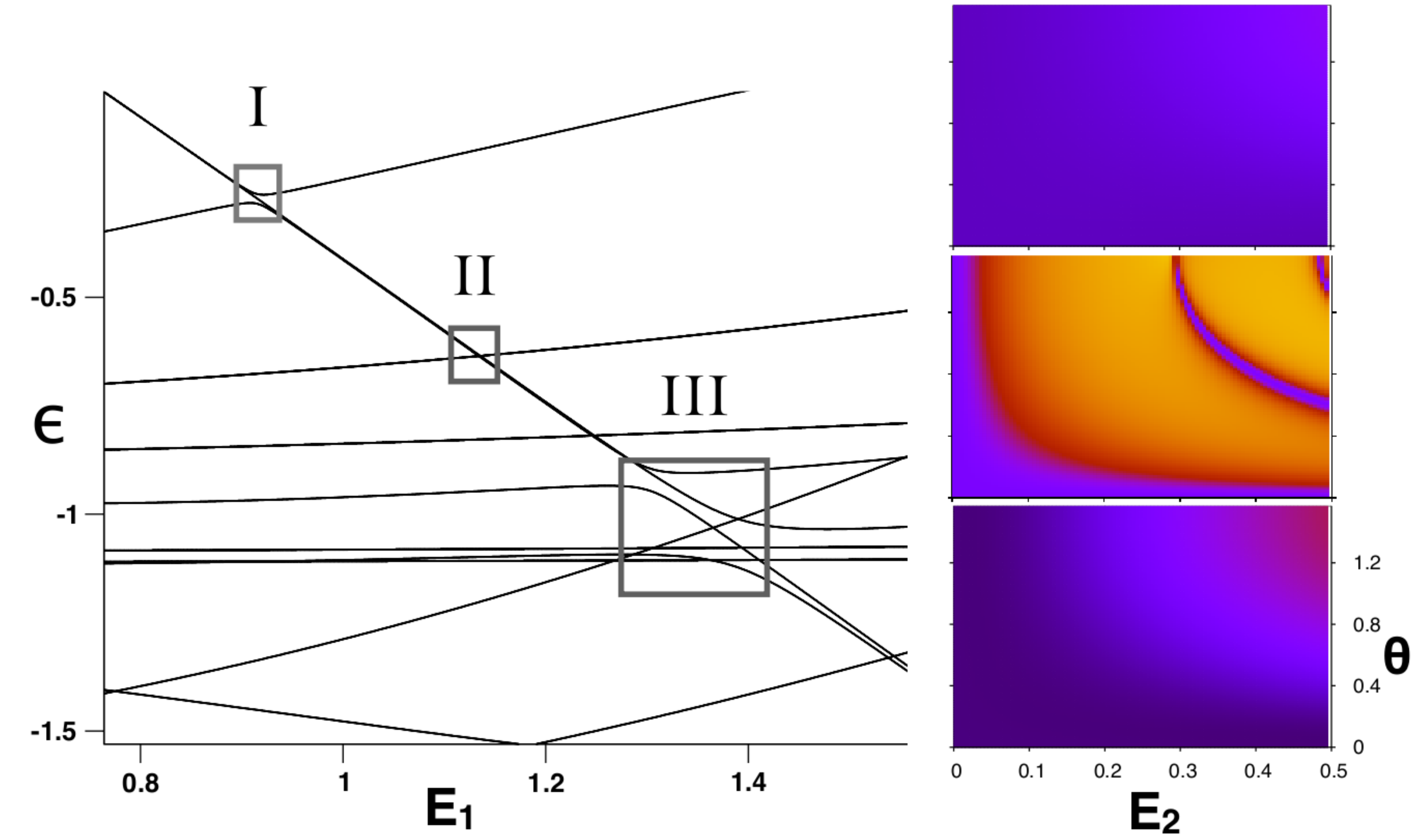}
\caption{Left panel shows a portion of the quasienergy spectrum as a function of the $\omega$-field amplitude $E_{1}$. The other fixed parameters are $E_{2}=0$, $\hbar = 0.2$ and $\omega = 2$. Three regions I, II, and III highlight different types of avoided crossings.  Right column shows the $(E_{2},\theta)$ control landscapes with fixed $E_{1}=0.919$ (bottom, case I), $E_{1}=1.129$ (middle, case II), and $E_{1}=1.399$ (top, case III). All landscapes in this plot are computed over an equally spaced $101 \times 101$ grid of the relevant variables.}
\label{fig3}
\end{center}
\end{figure}

In order to study the quantum-classical correspondence for the results shown in Fig.~\ref{fig2} and Fig.~\ref{fig3}  it is important to establish the phase space nature of the Floquet states that play an important role in the dynamics of the initial state of interest. Consequently, we compute the Husimi distribution\cite{husimi} 
\begin {equation}
\rho_{H}(q_{c},p_{c}) =\frac{1}{2\pi} \left | \langle  \Phi_{\alpha} |\psi   \rangle     \right |^{2},
\label{husimi}
\end{equation}
given by the overlap of coherent states $\psi(q;q_{c},p_{c})$ localized at various phase space points $(q_{c},p_{c})$ (cf. eq.~\ref{cohstate} ) with the Floquet states of interest $|\Phi_{\alpha}\rangle$.  Note that a scaling factor involving the Planck constant in the denominator of eq.~\ref{husimi} has been ignored. This is of no consequence to the present study where only the qualitative features of the Husimi distributions  are of interest.
The phase space nature of the various Floquet states thus obtained are expected to be important in understanding the features of the control landscapes.

\subsection{Role of the irregular Floquet states}

\begin{figure}
\begin{center}
\includegraphics[width=0.5\textwidth]{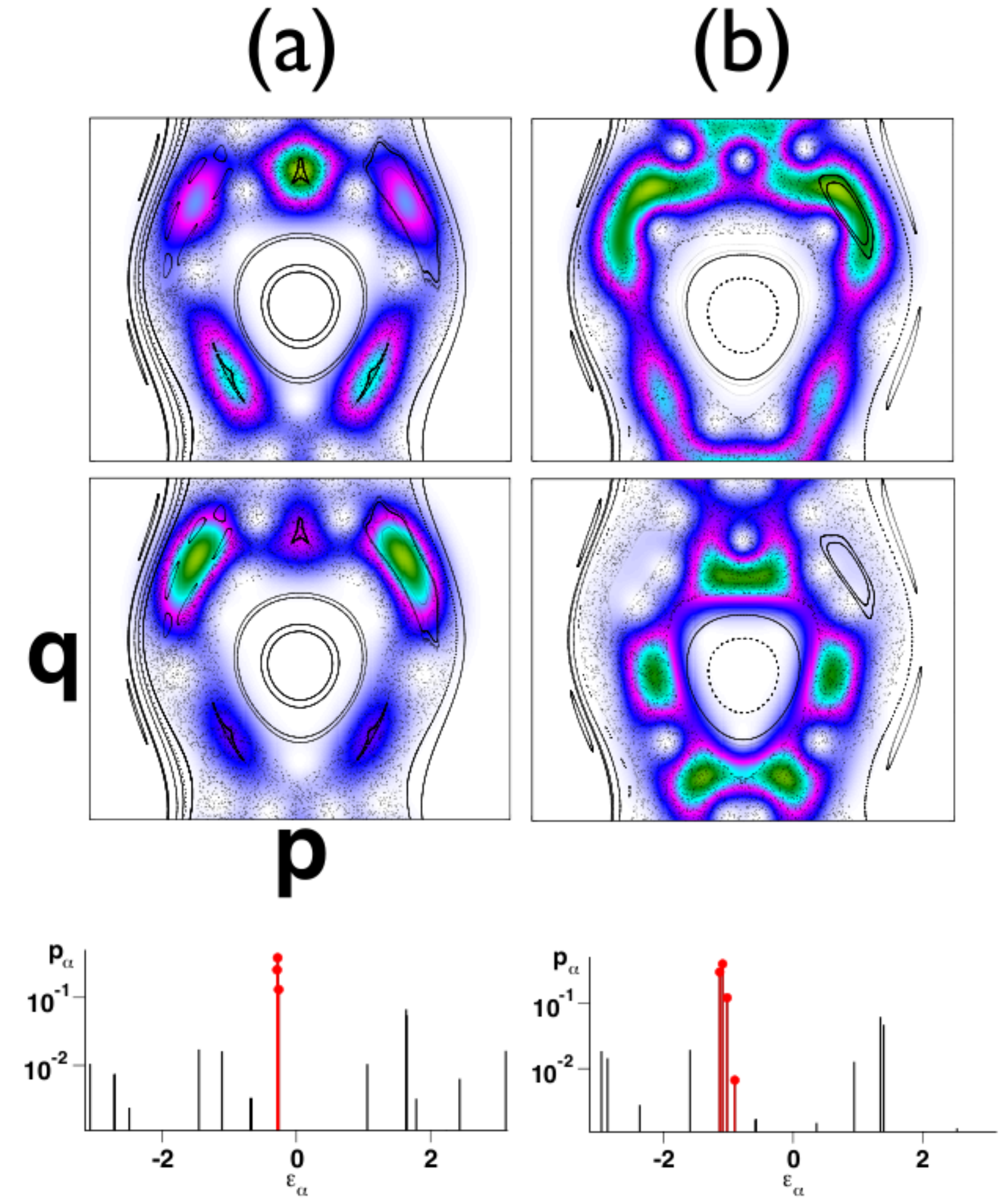}
\caption{Husimi distributions of delocalized Floquet states participating in the dynamics of a coherent state centered on the left $1:1$ resonance island for $E_{2}=0$. (a) Floquet states for $E_{1} \approx 0.919$ influenced by the $1:3$ resonance, corresponding to case $I$ in Fig.~\ref{fig1} and involved in an avoided crossing. (b) Two of the Floquet states for $E_{1} \approx 1.399$ delocalized in the chaotic region, corresponding to case $III$ in Fig.~\ref{fig1}. Axis range of the phase space are identical in all the plots and maximum Husimi density is in green. The overlap intensity spectrum in each case is also shown.}
\label{fig4}
\end{center}
\end{figure}

In Fig.~\ref{fig4}(a) the Husimi representation of two of the key Floquet states, corresponding to the case I of Fig.~\ref{fig3} with $E_{2}=0$, are shown along with the overlap intensity spectrum. Clearly, a state influenced by the $1:3$ nonlinear resonance is involved in the avoided crossing observed in Fig.~\ref{fig3} and this scenario, also highlighted in an earlier work\cite{latka94}, is typical of the RAT mechanism. The $1:3$ state continues to influence the initial state dynamics for increasing amplitude of the $2\omega$-field. Thus, the lack of bichromatic control seen in Fig.~\ref{fig3} (landscape, middle panel) arises due to the phenomenon of RAT and confirms earlier predictions made in the context of local control of the dissociation of a driven Morse oscillator\cite{sethi09}. 

\begin{figure}
\begin{center}
\includegraphics[width=0.45\textwidth]{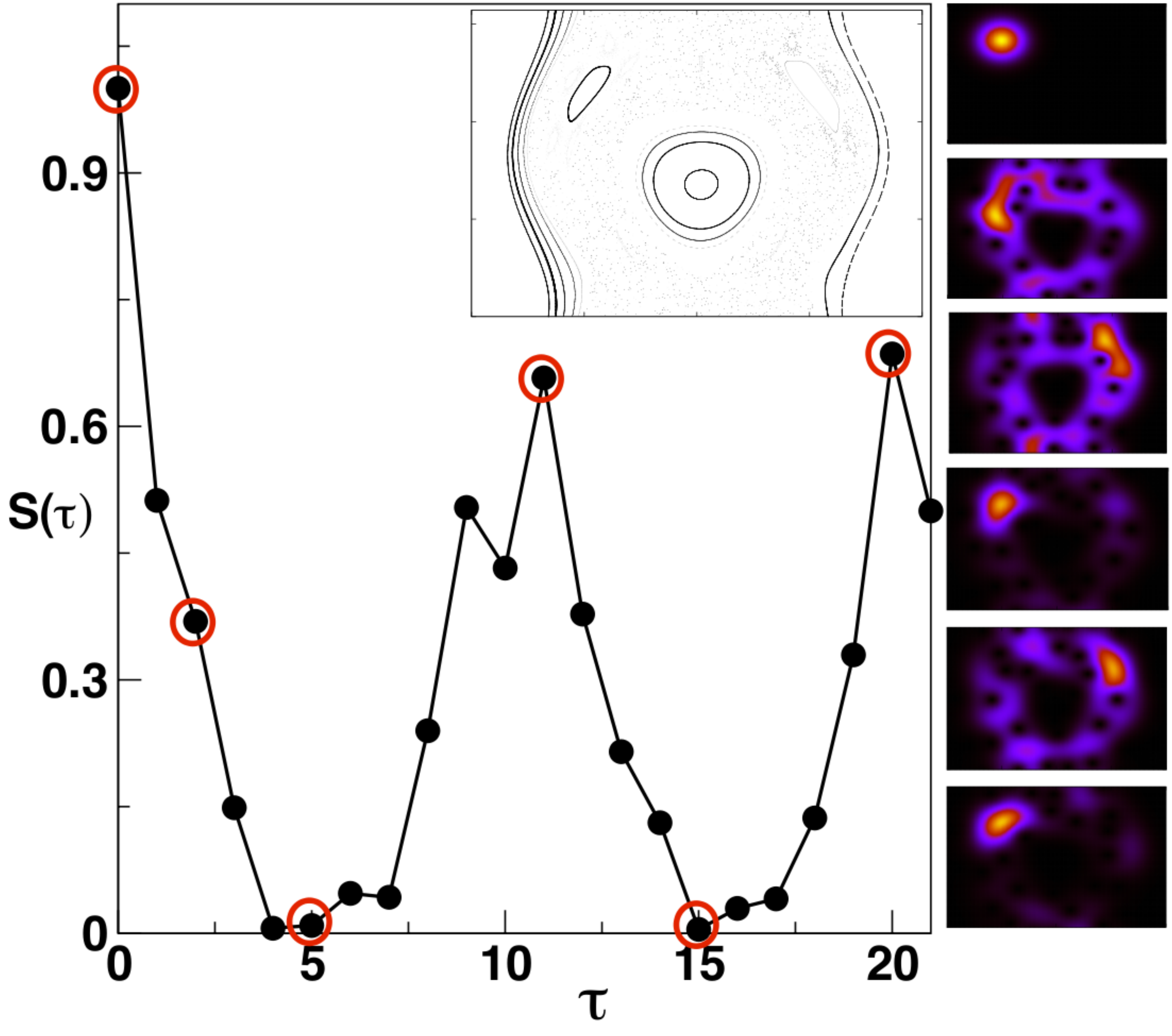}
\caption{Survival probability $S(\tau)$ for a symmetry broken case with $E_{1}=1.309$, $E_{2}=0.3$, $\hbar=0.2$, and $\theta=\pi/2$.  Initial state is centered around the left island in the classical phase space (shown in the inset) and time $\tau$ is in units of the field period. Dynamical tunneling is observed as shown by the snapshots of time evolving Husimi distributions in the left column. The Husimis are shown at select times, with the top most panel being at $\tau = 0$ and the bottom most panel at $\tau=20$, highlighted by red circles in the survival probability plot.}
\label{fig5}
\end{center}
\end{figure}

Results for the comparatively more complex case III are shown in Fig.~\ref{fig4}(b) and it is immediately clear that the dynamics involves at least four Floquet states with two of them being significantly delocalized in the phase space. Interestingly, of the two most delocalized Floquet states shown in Fig.~\ref{fig4}(b), one of them resembles a ``Janus" state\cite{mouchet06} which has its Husimi density localized around the border between the central stable island and the chaotic sea. Such states are expected\cite{mouchet06} to be involved in CAT and are less sensitive to the symmetry breaking induced by the $2\omega$-field. Indeed, our computations show that the delocalized states persist upon introducing the symmetry breaking field and continue to play an important role in the dynamics of the initial state. Thus, the ``wall of no control" seen in the $L(E_{1},E_{2},\pi/2)$ landscape  in Fig.~\ref{fig2} (top) can be associated with the participation of such  delocalized Floquet states. A similar observation was made in an earlier work\cite{sethi08} on bichromatic control of tunneling in the driven double well system. Note that in the context of directed transport case III corresponds to avoided crossing between states localized in the regular transporting islands and states delocalized in the chaotic sea. Further confirmation for the role of played by the delocalized Floquet states comes from Fig.~\ref{fig5} where snapshots of the time evolving Husimi distributions are shown. It is clear  that the survival probability oscillates with a time period of $\sim 10 \tau$ corresponding to significant revivals of the initial coherent state. Furthermore, the snapshots at specific times indicate the involvement of the delocalized Floquet states shown in Fig.~\ref{fig4}(b) and hence the lack of control observed in the landscape. 

\begin{figure}
\begin{center}
\includegraphics[width=0.45\textwidth]{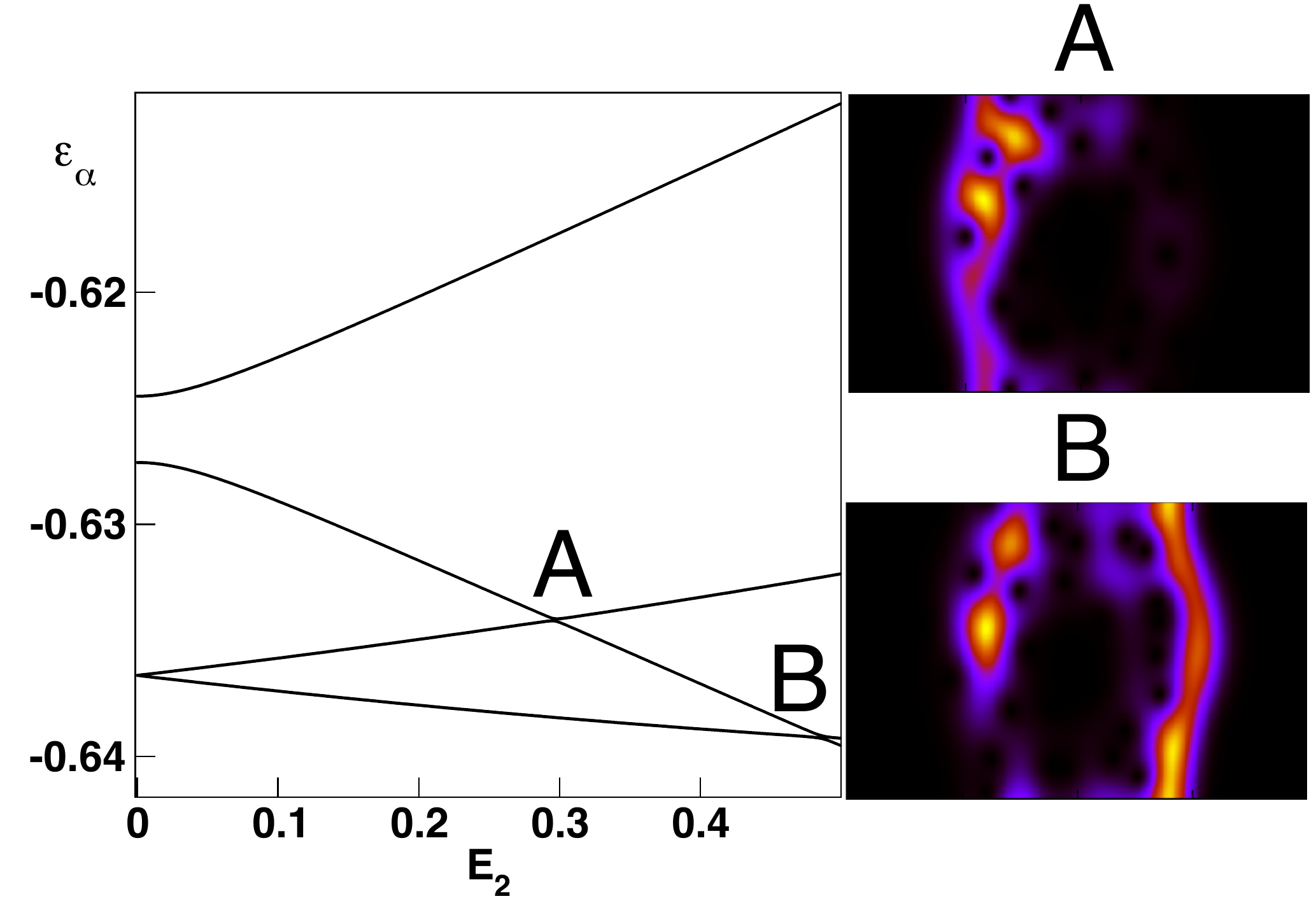}
\caption{Variation of the quasienergies with $E_{2}$ for fixed $E_{1} \approx 1.129$, $\hbar=0.2$, and $\theta=\pi/2$. Note that one of the Floquet states is involved in a sharp avoided crossings with desymmetrized rotor states at $E_{2} \approx 0.3$ and $0.5$, labeled as A and B respectively. Right column shows the snapshots of Husimi representation for the time evolved initial state  at $\tau=1146$ and $\tau=1454$ corresponding to A and B respectively. In both cases, the times (in units of field period) are chosen when the survival probability is a minimum.}
\label{fig6}
\end{center}
\end{figure}

In contrast to case I and III discussed above, Fig.~\ref{fig3} (middle) shows that in case II it is possible to control the dynamics. The key difference here is that the observed avoided crossing in Fig.~\ref{fig3} involves the regular tunneling doublet and a regular rotor doublet. Note that here too one observes lack of control in the $L(E_{1},E_{2},\pi/2)$ landscape in isolated regions around $E_{2} \approx 0.3$ and $E_{2} \approx 0.5$.  At these isolated points the dynamical tunneling process is certainly controlled by the $2\omega$-field but the initial coherent state does not localize. Instead, as shown in Fig.~\ref{fig6}, a two state sharp avoided crossing induced by the symmetry breaking field results in the delocalization of the initial state. Our computations show that very little to no amplitude is built up in the right $1:1$ island over long times and hence quite different from the results shown in Fig.~\ref{fig5} . Thus,  the mechanism for the lack of control is distinct from cases I and III. This is also reflected in Fig.~\ref{fig3} (middle) wherein small variations in the control knobs $E_{2}$ or $\theta$ disentangles the Floquet states and lead to control.

\subsection{Influence on directed transport}

It has been argued that symmetry breaking leads to directed transport classically as well as quantum mechanically. Classically, a non-zero dc-current $J_{\rm ch}$ comes about\cite{denisov01} due to the desymmetrization of the chaotic layer and perturbative arguments lead to $J_{\rm ch} \sim E_{1}^{2}E_{2} \sin \theta$. Quantum mechanically, however, significant enhancements in the current can occur when the relevant Floquet states are involved in an avoided crossing\cite{denisov07}. In particular, Denisov et al showed that avoided crossings between  Floquet states localized on the $1:1$ transporting island and  Floquet states delocalized in the chaotic layer play an important role. Moreover, tuning the $(E_{2},\theta)$ control knobs  allows one to vary the extent of the  enhanced current.

From the results in the previous section, as exemplified by Fig.~\ref{fig4}(b) and Fig.~\ref{fig5}, it is evident that Floquet states delocalized in the chaotic layer are also responsible for the loss of bichromatic control due to CAT. A question then arises - to what extent will the failure of symmetry breaking influence the directed transport? Is it possible for CAT to partially suppress the magnitude of the asymptotic current? To this end we compute the quantum asymptotic current
\begin{equation}
J(t_{0}) = \sum_{\alpha}  |C_{\alpha}(t_{0})|^{2} \langle p \rangle_{\alpha}
\label{qmcurrent}
\end{equation}
for an initial state which is an eigenstate of the momentum with eigenvalue zero. Note that the study of Denisov et al.\cite{denisov07} utilized such an initial state to highlight the role of the delocalized Floquet states in enhancing currents at resonance. In the above equation, $\langle p \rangle_{\alpha}$ denotes the average momentum in the Floquet state $|\Phi_{\alpha}\rangle$ and $C_{\alpha}(t_{0})$ represents the expansion coefficient of the initial state in the Floquet basis, with $t_{0}$ being the initial time.

\begin{figure}
\begin{center}
\includegraphics[width=0.48\textwidth]{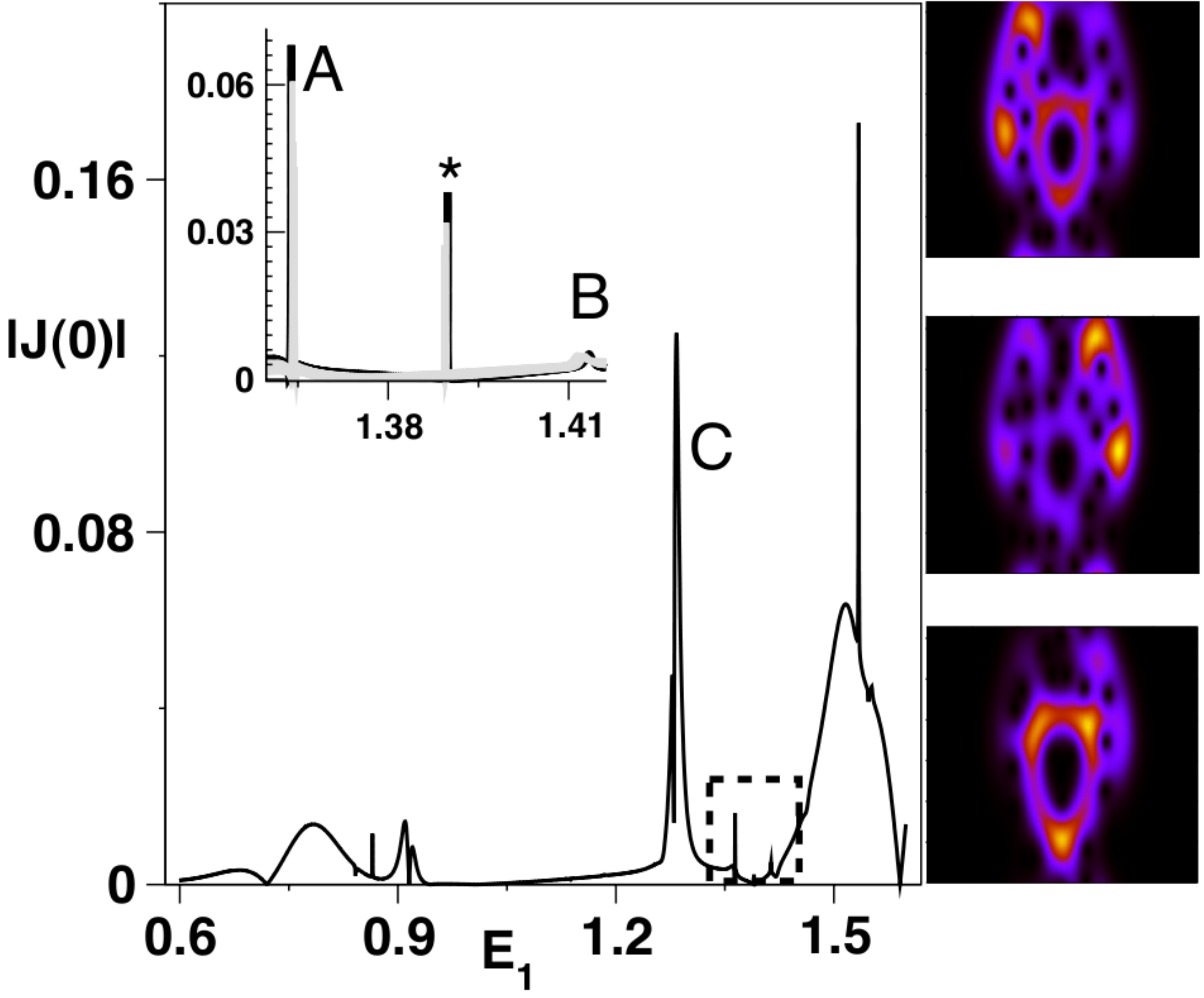}
\caption{Magnitude of the quantum asymptotic current given by eq.~\ref{qmcurrent} as a function of $E_{1}$ with $E_{2} = 0.3$,  $\theta=\pi/2$, and initial time $t_{0}=0$. The initial state is a plane wave state with zero wavevector. Inset shows the magnification of the region enclosed by the dashed rectangle. In the inset data for $\theta=\pi/3,\pi/4$, and $\pi/5$ with a finer resolution are also shown in gray. Right panels show the Husimi representation of selected Floquet states involved in avoided crossings for $\theta=\pi/2$. Top, middle and bottom panels correspond to A, B (inset) and C in the current plot. Axes, their range and color code are same as in the previous figures of Husimi representations. }
\label{fig7}
\end{center}
\end{figure}

In Fig.~\ref{fig7} the computed asymptotic current for maximum desymmetrization and initial time $t_{0}=0$ is shown. Note that, quantum mechanically and in contrast to the classical case, the asymptotic current is expected to be dependent on the initial time. However, as argued earlier\cite{denisov07}, the strong enhancements in $J(0)$ due to resonant interaction between Floquet states are independent of the initial time.  Several such enhancements can be seen in Fig.~\ref{fig7} and indeed do correspond to avoided crossings between Floquet states (localized and delocalized) in the central region of the phase space (cf. Fig.~\ref{fig1}) with the Floquet states localized in and around the $1:1$ transporting islands. Nevertheless, two key observations can be made. Firstly, $|J(0)|$ is significantly smaller in the range $E_{1} \in (1.30,1.45)$,  precisely corresponding to the region III in the control landscape shown in Fig.~\ref{fig3}, with a few fluctuations. Secondly, as shown in the inset to Fig.~\ref{fig7}, the value of current in this region is rather insensitive to the value of the relative phase. Note that much larger enhancements, for instance case C in Fig.~\ref{fig7}, are seen just outside this range. In order to gain further insights the phase space nature of the Floquet states that are involved in avoided crossings in this region were studied. In Fig.~\ref{fig7} three such states are shown with cases A and B being in region III and case C just outside this region. It is immediately clear that case C shows strong asymmetry when compared to cases A and B. Based on our understanding of the control landscape it seems plausible that CAT in region III is responsible for small values of the current and, more importantly, a somewhat suppressed enhancement despite the presence of resonantly interacting Floquet states.

\section{Concluding remarks}

In this work we have shown that bichromatic control of dynamical tunneling can be compromised due to the presence of delocalized Floquet states in the classical phase space.  In essence, participation of the  Floquet states delocalized in the chaotic regions of the phase space results in chaos-assisted tunneling which is robust despite breaking of the symmetries upon addition of a second field. Since the present work deals with the control of dynamical tunneling, a quantum process with no classical limit, the results do not necessarily invalidate the idea of laser induced symmetry breaking approach to control. Indeed, our landscape computations for decreasing values of the Planck constant do show that the ``pillars" of no bichromatic control are reduced to ``dust" in the deep semiclassical limit. Nevertheless, it remains to be seen as to whether chaos-assisted tunneling  can reduce the efficiency or significantly modify the mechanism of control based on the principle of interference of optically induced pathways\cite{franco10}. 

Moreover, our analysis suggests that such a loss of bichromatic control can potentially influence current rectification in driven systems. An earlier study\cite{gong04} by Gong and Brumer has also hinted at the possibility of reduced directed currents in Hamiltonian ratchets due to dynamical tunneling between a chaotic sea and and transporting islands embedded in the chaotic sea. However, further  detailed studies are required  in order to clearly establish the role of chaos-assisted tunneling in ratcheting systems with\cite{kato13} and without dissipation. In the latter case, the role of hierarchical states\cite{ketzmerick00,backer02} in both control of dynamical tunneling and generation of directed transport needs to be explored further. Note that one possible explanation for the ``pillars" of no control being reduced to ``dust" in the $\hbar \rightarrow 0$ limit may be related to the fact that the fraction of hierarchical states goes to zero in this limit.

As a final note we point to an interesting parallel\cite{ivanov12} between bichromatically driven multilevel systems and the problem of molecular alignment and orientation on a plane. Assuming that the present analysis can be extended to this observation, our work hints at the possibility that  for certain values of the orienting DC field and a range of aligning field strengths, irrespective of the angle between the two fields, coherent control should be ineffective.

\acknowledgments

AS is grateful to CSIR, India for a doctoral fellowship.

\end{document}